\documentclass[%
aip,
rsi,%
amsmath,amssymb,
reprint,%
]{revtex4-1}

\usepackage{graphicx}
\usepackage{dcolumn}
\usepackage{bm}

\begin{document}
	
	\title{Tutorial: Photonic Neural Networks in Delay Systems}
		
		\author{D. Brunner}
		\email{daniel.brunner@femto-st.fr}
		\affiliation{FEMTO-ST Institute/Optics Department,CNRS \& Univ. Bourgogne Franche-Comt\'{e} CNRS, 15B avenue des Montboucons, 25030 Besan\c{c}on Cedex, France}
		
		\author{B. Penkovsky}
		\affiliation{FEMTO-ST Institute/Optics Department,CNRS \& Univ. Bourgogne Franche-Comt\'{e} CNRS, 15B avenue des Montboucons, 25030 Besan\c{c}on Cedex, France}
		
		\author{B. A. Marquez}
		\affiliation{FEMTO-ST Institute/Optics Department,CNRS \& Univ. Bourgogne Franche-Comt\'{e} CNRS, 15B avenue des Montboucons, 25030 Besan\c{c}on Cedex, France}
		
		\author{M. Jaquot}
		\affiliation{FEMTO-ST Institute/Optics Department,CNRS \& Univ. Bourgogne Franche-Comt\'{e} CNRS, 15B avenue des Montboucons, 25030 Besan\c{c}on Cedex, France}
		
		\author{I. Fischer}
		\affiliation{Instituto de F\'{\i}sica Interdisciplinar y Sistemas Complejos, IFISC (UIB-CSIC), Campus Universitat de les Illes Balears, E-07122 Palma de Mallorca, Spain.}
		
		\author{L. Larger}
		\affiliation{FEMTO-ST Institute/Optics Department,CNRS \& Univ. Bourgogne Franche-Comt\'{e} CNRS, 15B avenue des Montboucons, 25030 Besan\c{c}on Cedex, France}

		\date{\today}
		
		\begin{abstract}
			
			Photonic delay systems have revolutionized the hardware implementation of Recurrent Neural Networks and Reservoir Computing in particular.
			The fundamental principles of Reservoir Computing strongly benefit a realization in such complex analog systems.
			Especially delay systems, potentially providing large numbers of degrees of freedom even in simple architectures, can efficiently be exploited for information processing.
			The numerous demonstrations of their performance led to a revival of photonic Artificial Neural Network.
			Today, an astonishing variety of physical substrates, implementation techniques as well as network architectures based on this approach have been successfully employed.
			Important fundamental aspects of analog hardware Artificial Neural Networks have been investigated, and multiple high-performance applications have been demonstrated.
			Here, we introduce and explain the most relevant aspects of Artificial Neural Networks and delay systems, the seminal experimental demonstrations of Reservoir Computing in photonic delay systems, plus the most recent and advanced realizations.
			
		\end{abstract}
\maketitle

\section{Introduction}

Artificial Neural Networks (ANNs) are based on computational concepts fundamentally different from the current computational workhorse, the Turing-von Neumann concept.
Inspired by a strongly simplified interpretation of the human brain's structure, large numbers of simple nonlinear elements (neurons) are connected (synaptic links) into large networks.
Information processing in ANNs usually relies on numerous simple nonlinear transformations and large scale linear matrix multiplications.
The implementation of these operations in von Neumann architectures is highly inefficient, as it requires massive parallelism.
Originally, ANN concepts were already introduced more than five decades ago \cite{McCulloch1943,Werbos1974}.  
Yet, the inefficiency of their emulation strongly limited their exploitation; not for lack of powerful concepts, interest, potential or applications, but for lack of adequate computing power and computing architectures.
During the past decade, this has changed significantly, and currently ANNs are widely considered key to future technological advance.
Nowadays, tasks reaching astonishing levels of complexity and abstraction can be solved with high accuracy:  ANNs can describe images \cite{LeCun2015}, identify human faces \cite{amos2016openface} as well as spoken digits \cite{Graves2013a}.

The recent breakthroughs have largely profited from two strategic advances: the availability of large amounts of data, and of economic high-performance computing devices.
Optimizing ANNs to solve specific tasks requires the adjustment of their connectivity structure, in most cases a slow and painstaking process. 
As current ANN algorithms chiefly rely on large sets of example data for this \textit{training} or \textit{learning} process, the surge of data created by modern communication technology, i.e. the Internet, plays a significant role in recent breakthroughs in training complex ANNs.
A second major driving force is the off-the-shelf availability of high-performance computing devices like field programmable gate array (FPGAs) and graphics processing units (GPUs).
These are better suited to implement ANNs' massive parallelism than standard central processing units.

By now, the potential and importance of ANNs is widely recognized.
Yet, even the recent astonishing developments cannot mask the fact that currently no ideal ANN-specific hardware, which fully implements physical hardware neurons and physical \textit{synaptic links}, exists.
With such a novel platform, many orders of magnitude could be gained in speed and energy efficiency.
At this place the human brain with its learning and information processing capabilities, while only consuming about 25W of power, serves as a humbling benchmark, reminding us of what \textit{is} achievable.
The quest for new hardware substrates and hardware enabling ANN concepts is gaining considerable momentum, but it remains a field where much still awaits its discovery. 

The Reservoir Computing (RC) concept is of special relevance to the implementation of ANNs in unconventional physical substrates.
The concept was developed in parallel by multiple groups \cite{Jaeger2001,Maass2002,Jaeger2004,Steil2004}.
Reservoir Computers offer a compromise between performance and an implementation-friendly ANN topology.
Starting in 2011, electronic nonlinear delay systems \cite{Appeltant2011}, in 2012 \cite{Larger2012,Paquot2012} opto-electronic, soon followed by all-optical nonlinear delay systems \cite{Duport2012,Brunner2013a} demonstrated fully implemented analog reservoirs.
These hardware reservoirs with a ring-like delay topology allowed for the first time to physically implement analog ANNs consisting of 100s or 1000s of neurons based on practical experimental setups.
Meanwhile, numerous demonstrators and even fully autonomous and fully operational RCs \cite{Antonik2016} have been implemented based on this principle.
Finally, further conceptional simplifications of such systems allowing for high-speed channel equalization in optical communication systems \cite{Argyris2018}, extended multi-delay architectures \cite{Martinenghi2012}, as well as modifications of the training process of the ANN including the input connectivity\cite{Hermans2015a} have been introduced and demonstrated.

A recent review \cite{VanderSande2017} provides an overview of photonic RC in general.
In this tutorial we will focus on concepts, techniques, possibilities and limitations of RC based on photonic nonlinear delay systems.
However, general aspects of the implementation and different concepts are mostly transferable to other, non-photonic substrates.
First, we will introduce fundamental aspects and properties of ANNs. From there we will move on to discuss their implementation in nonlinear delay systems.
Then, we will present multiple seminal experiments which implemented these concepts in different nonlinear delay systems, provided key insight into aspects of particular relevance to hardware realizations, or further advanced the concept.

\subsection{Artificial Neural Networks}

A neuron typically consists of a body (soma), dendrites receiving electrical inputs, and a long axon sending electrical spikes towards other neurons.
Nevertheless, the dynamics of a neuron can be very intricate, which makes their implementation computationally expensive.
Emulating or physically implementing an ANN with biologically plausible details can therefore be a daunting task, and here we are chiefly interested in their computational capacity.
Instead of trying to veritably reproduce detailed electrodynamic properties of the neuron, many aspects, including its excitable properties giving rise to time-dependent spiking dynamics, are abstracted away for the sake of implementation efficiency.
It is usually sufficient to describe only the connectivity properties of a nonlinear network-node as a computational unit.
Such an artificial neuron is called a \textit{perceptron} and is expressed via the following equation: 
\begin{equation}
y=f\left(\sum_{i=0}^{M}w_{i}u_{i}\right),\label{eq:perceptron}
\end{equation}
\noindent where the dot product $\sum_{i=0}^{M}w_{i}u_{i}$ represents information inputs, i.e. the analogy to dendrites; $f(\cdot)$ is the \textit{activation} function, typically a monotonic nonlinear function representing the neural-cell's nonlinear transformation; and $y$ is the perceptron's output, similar to the axon's outward connection.
Sometimes Eq. \eqref{eq:perceptron} is written including a bias $w_0$, which is equivalent to having an additional constant input $u_0 = 1$ in our notation.

\begin{figure}[!t]
	\begin{centering}
		\includegraphics[width=6cm]{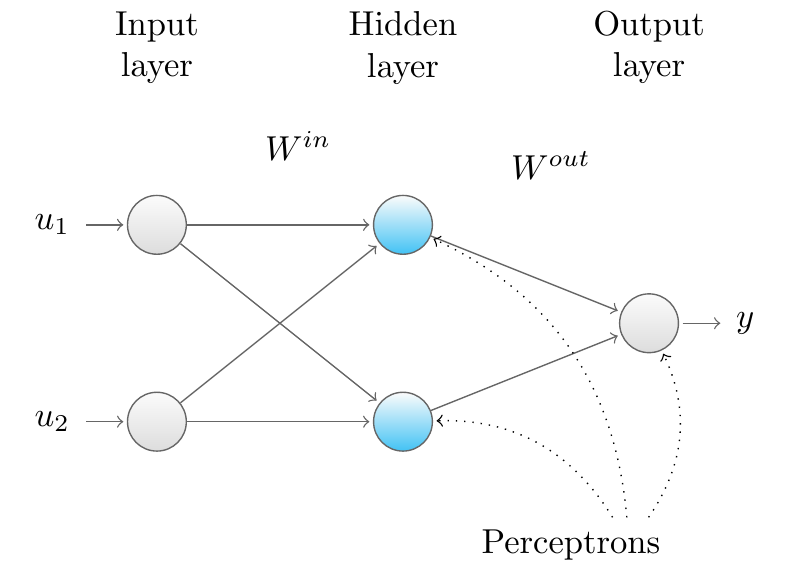}
		\par\end{centering}
	\caption{A feedforward neural network (FNN) consisting of three layers: an input layer $\mathbf{u} = (u_1, u_2)$, the hidden layer consisting of perceptrons (blue circles), each described by Eq. \eqref{eq:perceptron}.
		From there information is passed to output layer $\mathbf{y}$, which is yet another perceptron with inputs $\mathbf{x}$.
		Hidden and output layer connection weights are given by matrices $W^{in}$ and $W^{out}$, respectively.}
	\label{fig:feedforward-nn}
\end{figure}

To perform computation, such perceptrons are connected with each other, thereby forming networks.
In Fig. \ref{fig:feedforward-nn} we schematically illustrated a typical ANN.
Due to the unidirectional network coupling, i.e. from left to right, the network is called a \emph{Feedforward} Neural Network (FNN).
Hence, the network is expressed by the following system of equations:
\begin{equation}
\begin{array}{lll}
\mathbf{x} & = & f_{1}\left(W^{in}\mathbf{u}\right),\\
\mathbf{y} & = & f_{2}\left(W^{out}\mathbf{x}\right),
\end{array}
\end{equation}
\noindent where vector $\mathbf{u}$ is the input information, matrices $W^{in}$ and $W^{out}$ are connection weights,
$f_{1}$ and $f_{2}$ are activation functions, vector $\mathbf{x}$ is the internal (hidden layer) network state and vector $\mathbf{y}$ is the computed result.
Note that the network is not unique, i.e. there exist many possible configurations of $W^{in}$ and $W^{out}$ solving the same problem.
A common problem is therefore finding the best architecture suitable to solve a particular problem.
Nowadays, networks with more than one hidden layer are used in so-called \textit{deep learning} architectures \cite{Goodfellow2016}.
Such networks are typically trained using the \textit{backpropagation} algorithm \cite{Rumelhart1986b,Rumelhart1986}, in which the weights are updated in inverse order to the processing direction, i.e. starting from the last and moving to the first layer.

\begin{figure}[!t]
	\begin{centering}
		\includegraphics[width=6cm]{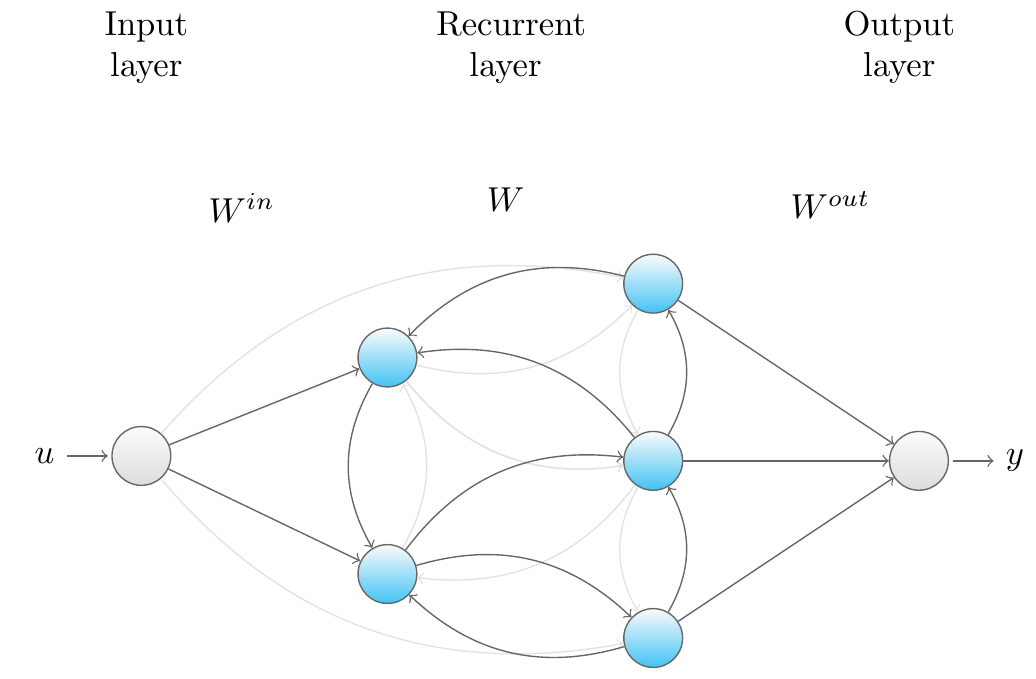}
		\par\end{centering}
	\caption{Schematic of information processing flow in a \emph{Recurrent} Neural Network (RNN).
		Other than for FFNs, connections of the hidden (recurrent) layer link to elements of the same layer and even themselves, introducing recurrence.
		The recurrent layer therefore can exhibit dynamics including echoes from previous inputs, and RNNs are therefore capable to process information within a temporal context.}
	\label{fig:recurrent-neural-network}
\end{figure}

A second important class of ANNs are \emph{Recurrent} Neural Networks (RNNs), as schematically illustrated in Fig. \ref{fig:recurrent-neural-network}.
The essential difference between FNNs and RNNs is the connectivity structure of their hidden layer.
In RNNs, the hidden layer includes connections between individual neurons.
Flow of information is therefore not unidirectional, i.e. information can circle within the hidden layer due to these \emph{recurrent} connections.
The hidden layer is therefore typically referred to as the \emph{recurrent layer}.
Due to these non-exclusively forward-directed connections in Fig. \ref{fig:recurrent-neural-network}, the recurrent layer includes internal connections which propagate information along temporal loops.
While FNNs are memoryless and can be understood as universal function approximators, RNNs can be regarded as \emph{algorithms}.
The main difference is that functions (or maps) do not have memory (they are \emph{stateless}), while algorithms do.
It has been shown that RNNs are universal approximators of dynamical systems \cite{Funahashi1993} and, moreover, that theoretically they are Turing equivalent \cite{Kilian1996,Grigoryeva2018}. 
Finally, because of the presence of recurrent loops, RNNs exhibit some analogies with aspects of the biological brain \cite{Douglas2004}.

Mathematically, a RNN's recurrent layer can be expressed as:
\begin{equation}
\mathbf{x}(n+1)=f\left(W^{in}\mathbf{u}(n+1)+W\mathbf{x}(n)\right),\ n=1,2,\ldots,T,\label{eq:RNN-state}
\end{equation}
\noindent where matrix $W^{in}\in\mathbb{R}^{N\times M}$ is the input map, $W\in\mathbb{R}^{N\times N}$ is the map of the previous state of the recurrent layer, and $\mathbf{x}(n)$ is the network's \emph{internal state}.
The recurrent layer consist of $N$ neurons, and the input information contains $M$ elements at each instance of integer time $n$.
The result of computation is then to be realized utilizing the transformations provided by $\mathbf{x}(n, \mathbf{u})$.

\subsection{Training a Reservoir readout layer}
\label{sec:RC_training}

Using the \textit{unfolding in time} procedure, it can be shown that RNNs are equivalent to deep FNNs with, at least theoretically, an infinite number of hidden layers.
Although the back-propagation algorithm can therefore be adapted to RNN training, it becomes challenging to achieve training convergence as the large number of layers often results in exploding or vanishing gradients \cite{Pascanu2012}.
In RC, instead of modifying all the RNN layers, only the final \textit{readout} layer is optimized during training, keeping the (often randomly constructed) recurrent layer's connectivity $W$ unmodified during the training procedure.
As a result, training convergence is efficiently achieved.
The linear readout layer $W^{out}$ (Fig. \ref{fig:recurrent-neural-network}) is expressed as:
\begin{equation}
\mathbf{y}(n)=W^{out}\mathbf{x}(n), \label{eq:RC-training}
\end{equation}
\noindent where matrix $W^{out}\in\mathbb{R}^{K\times N}$ is a readout map; $K$ is the output dimensionality, i.e. the number of parallel results provided by $\mathbf{y}(n)$.

The linear readout weights $\mathbf{W}^{out}$ are obtained from processed data samples (Eq.~\eqref{eq:RNN-state}), for example using the following ridge regression:
\begin{equation}\label{eq:ridge-regression}
\mathbf{W}^{out}=(\mathbf{M}_x\cdot \mathbf{M}_x^{\text{T}} + \lambda\cdot\mathbf{I} )^{-1} (\mathbf{M}_x\cdot \mathbf{T}^{\text{T}}),
\end{equation}
\noindent where $\lambda \ll 1$ is a small regularization constant.
$\mathbf{M}_x \in \mathbb{R}^{N \times Q}$ is a feature matrix of horizontally concatenated state vectors $\mathbf{x}(n)$, $\mathbf{T} \in \mathbb{R}^{K \times Q}$ is the teacher matrix, which corresponds to the desired optimal computational results provided by the RC.
$Q$ denotes the number of training feature vectors typically identical to the number of input examples $T$.
As for the Reservoir feature matrix $\mathbf{M}_x$, $\mathbf{T}$ consists of horizontally concatenated state vectors of the output target $\mathbf{y}^{T}(n)$.
Training according to Eq. \eqref{eq:ridge-regression} requires targets $\mathbf{y}^{T}(n)$ to be known in advance, hence the scheme corresponds to supervised learning.

For classification tasks, the teacher is a \textit{one-hot encoded} matrix, i.e. consists of target answer vectors $\mathbf{y}^{T} \in \mathbb{R}^{K \times 1}$ where the only non-zero elements correspond to the correct class labels.
For prediction tasks, the teacher is the future value of the signal to be predicted from the input data $\mathbf{u}(n)$.
Due to the sensitivity of Reservoir state $\mathbf{x}(n)$ to previously injected information, trained output $\mathbf{y}(n)$ is capable of addressing such temporal problems.

\subsection{Implementation in nonlinear photonic systems}

The very architecture of ANNs was quickly identified as a major implementation or emulation bottleneck.
The massive parallelism required for an efficient realization of the large-scale matrix products, see Eqs. (\ref{eq:perceptron}-\ref{eq:RC-training}), cannot be mapped directly onto the von Neumann architecture.
Improvements can be achieved by utilizing GPUs and google's TPU \cite{Jouppi2017}, yet one can also approach the problem from a more fundamental point of view.
Other than electrons, photons are information carriers which do not interact for low intensities and in the absence of an interaction-mediating medium.
The following discussion's validity is therefore restricted to low optical intensities.

The spatial Fourier transforming property of a simple optical lens is an excellent illustration: under specific conditions, an entire image's Fourier-spectrum, including all spatial-frequencies supported by the lens's impulse response function, is simultaneously provided at its back focal plane \cite{Goodman2005}.
The information contained within an image is therefore processed in parallel, for example allowing convolving two independent spatial distributions fully in parallel \cite{Weaver1966}.
Electrons, however, do interact strongly, for example due to their Coulomb interaction.
Information encoded in the fundamental state of various electrons would therefore be modified in a nontrivial fashion by their nonlinear interaction, and parallel information-transmission along the same channel is frustrated.

Another advantageous property of photons is their inherent propagation at the speed of light, only influenced by the propagation medium's refractive index.
In particular, a photon's propagation speed does not depend on the length of a signal transmission line.
This is significantly different for signal transmission using electrons.
The response time of an electronic transmission line is given by its $RC$ constant, where $R$ and $C$ are Ohmic resistance and capacity, respectively.
In the leading order, an electric transmission line's RC constant linearly scales with its length.
This does not only induce delays, but also results in a convolution of the original information by the transmission line's impulse response function.
Increasing the transmission line's length does therefore inevitably reduce the maximum bandwidth it supports.

These aspects are of particular importance for the implementation of ANNs.
Implementations in spatio-temporal systems require a large number of connections, and their transmission line induced bandwidth limitations cannot easily be overcome by feature size reduction \cite{Smith1985}.
Also, in electronics one cannot easily leverage tools for increased parallelism such as wavelength or space  \cite{Richardson2013} division multiplexing, well established photonic techniques which for the case of wavelength multiplexing already are exploited for photonic ANNs \cite{Tait2014}.
Implementations in electronic delay systems suffer even stronger.
One can easily increase photonic delays and thereby increase the size of a photonic delay ANN without much influence on the system besides the delay time.
In electronics, long delay lines consisting of simple wire-connections are effectively ruled out for the introduced bandwidth limitations.
One typically has to fall back to digital implementations such a first-in-first-out memories (FIFO).

\subsection{Delay systems for Artificial Neural Networks}
\label{sec:DelayANNs}

Until this point, we have introduced ANNs as complex and high dimensional (Eq. \eqref{eq:perceptron}) and spatio-temporal systems (Eq. \eqref{eq:RNN-state}).
Their dimensionality is related to the number of independent neurons or nodes that a particular network architecture contains.
Crucially, comparable high dimensionality and complexity can be found in purely temporal oscillators such as dynamical systems with delayed feedback.
For the case of continuous-time oscillators, they can be mathematically described by delay differential equations (DDEs).
In a simple form, DDEs with a single delay are given by
\begin{equation}
\dot{\mathbf{x}}(t) = \mathbf{f}(t, \mathbf{x}(t), \mathbf{x}(t-\tau_D)), \\ \label{eq:dde_diff}
\end{equation}
\noindent where $\tau_D$ is the delay time resulting in the delayed feedback signal $\mathbf{x}(t-\tau_D)$ and $\mathbf{f}(\dots)$ a potentially multi-dimensional nonlinear transformation.
A classical example of Eq. \eqref{eq:dde_diff} are one-dimensional nonlinear delay system, which can be described by
\begin{equation} 
T_R \dot{x}(t) + x(t) = f[ x(t), \beta x(t-\tau_D), \rho u(t), \textbf{b}]. \label{eq:dde_num}
\end{equation}
\noindent This particular DDE has already been adjusted to be similar to the equational systems typically describing RNNs, see Eq. \eqref{eq:RNN-state}.
In Eq. \eqref{eq:dde_num}, nonlinearity $f(\dots)$ is only one-dimensional, $T_{R}$ is the system's response time and $\beta$ is the feedback gain.
Furthermore, Eq. \eqref{eq:dde_num} is extended by adding an external input $u(t)$, linearly scaled by $\rho$.
Finally, $\mathbf{b}$ is a set of constant parameters.

\begin{figure}[!t]
	\begin{centering}
		\includegraphics[width=6cm]{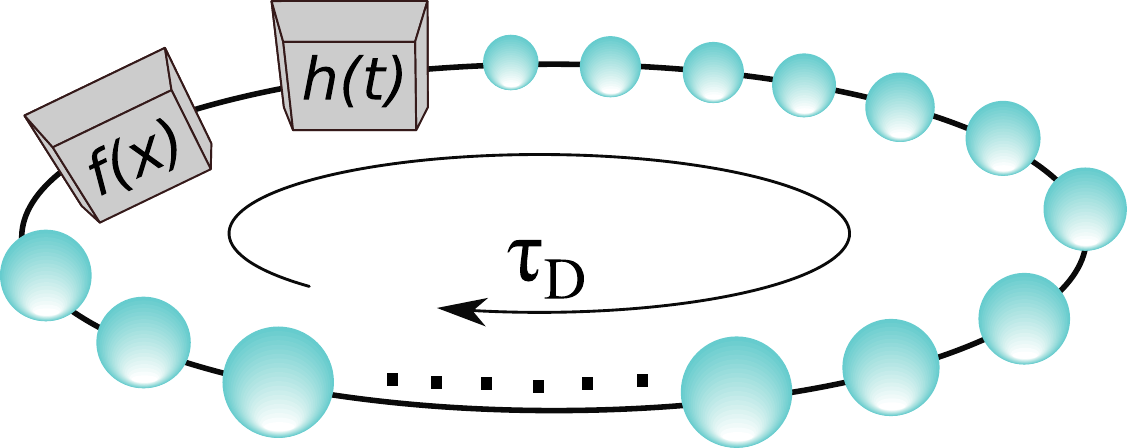}
		\par\end{centering}
	\caption{A DDE as a convolution product between an impulse response function $h(t)$ and a nonlinear function $f(x)$ having the delayed variable $x(t-\tau_D)$ as its argument.}
	\label{fig:circular}
\end{figure}

A DDE is an ordinary differential equation (ODE) modified through an additional input: its state-variable which is temporally shifted by time delay $\tau_{D}$.
While a seemingly small modification to a standard ODE, the delayed term $x(t-\tau_D)$ strongly modifies the system's dimensionality.
Due to this simple delay term, initializing the system requires defining its complete state during one delay interval $\tau_D$.
As we are dealing with a temporally continuous signal, in principle one requires an infinite amount of initial conditions, and mathematically speaking DDEs are equivalent to an infinite-dimensional system of ODEs \cite{Demidenko2009}.
Technically, the time-continuous initial state can be approximated by a discrete signal sampled at least twice faster than response time $T_R$.
Since typically $T_r\ll\tau_D$, DDE's are system's of sufficiently high dimensionality to implement the internal states of ANNs while consisting of only one single nonlinear element.
The dynamical evolution of a DDE can be numerically obtained with standard numeric integration techniques such as the Euler \cite{Sprott2007} or the Runge-Kutta 4th order method.

Experimental realizations of delay-dynamical systems \cite{Cornelles2015} have been implemented in many fields of science and technology, including electronic \cite{Appeltant2011}, electro-optic (EO) \cite{Neyer1982, Larger1, Chembo, Callan2010} and all-optical systems \cite{Soriano2013a}.
Employing such systems, high performance broadband chaotic communications~\cite{Argyris,larger2}, optical low-coherence sources \cite{Peil2006}, ultra-stable microwave sources \cite{Yao1994, Chembo2009, Maleki2011}, low-linewidth optical sources \cite{Brunner2017} and random number generation \cite{Uchida2008,Fang2014} have been demonstrated.
The fundamental properties of such oscillators have also received significant attention \cite{Cohen2008, Murphy2010, Ravoori2011, Illing2011, Illing2011b,Brunner2015, McNamara1988}.
Interestingly, these different applications require the delay oscillator to be operated in different regimes.
Chaos communication, random number generation and low-coherence emission rely on chaotic dynamics, for ultra high purity microwaves generation the delay systems are operated in the periodic, and for the low-linewidth optical sources in the stable regime.
The asymptotic steady state (equilibrium) regime is also most suited for information processing based on these systems \cite{Appeltant2011, Larger2012,Martinenghi2012}.
Operation beyond this fixed-point typically results in performance deterioration: the system's rendered inconsitent, as autonomous dynamics potentially result in different responses to identical input data \cite{Uchida2004}.
For the case of autonomous periodic orbits simple mechanisms can compensate some of their negative impact \cite{Marquez2018}.
But even for chaotic operation, consistent responses to additional input drives, as a necessary condition for information processing, can be achieved \cite{Oliver2016}.

\subsubsection{Spatio-temporal analogy}
\label{sec:DelayANNs_SpaceTempAnalogy}

Conventional neural networks are based on equations of coupled nonlinear elements.
It might not be directly apparent how time-continuous delay systems, as introduced by Eq. \eqref{eq:dde_num}, correspond to such a network comprising of discrete elements.
However, a direct spatio-temporal analogy between such networks and DDEs exists, illustrating that DDEs can be seen as a particular class of RNNs.
The analogy was first suggested in \cite{Arecchi1992} and later formally shown to be equivalent to a ring network topology in \cite{Larger2015b}.
In Fig. \ref{fig:circular}, such a network-representation of a delay system is schematically illustrated, and we will show that transforming a DDE into such a system only requires the system's nonlinearity $f(\cdot)$ and its impulse response function $h(t)$.
Considering the following DDE based on Eq. \eqref{eq:dde_diff}
\begin{equation} 
T_R \dot{x}(t) + x(t) = f \left(\beta x(t-\tau_D) + \rho u(t) + \Phi_0 \right), \label{eq:dde_ikeda}
\end{equation}
\noindent please note that in contrast to classical ANN models, we do not restrict $f(\cdot)$ the family of monotonous functions.
Often, a sinusoidal activation function is employed \cite{Larger2012,Larger2017}, for which Eq. \eqref{eq:dde_ikeda} corresponds to the Ikeda delay equation.
In this particular class of DDEs prominent in Optics, $f(\cdot)$ is easily obtained through a phase-to-intensity nonlinear transformation including optical interference between two fields.
Parameter $\Phi_0$ is a constant phase offset, other parameters are as previously introduced.

We will now discuss in what sense state $x(t)$ located within one $\tau_D$ can be understood as state of a virtual network node.
For convenience, we first normalize Eq. \eqref{eq:dde_ikeda} by introducing dimensionless entities $s := t / \tau_D$ and $\varepsilon := T_R / \tau_D$, resulting in
\begin{equation} 
\varepsilon \dot{x}(s) + x(s) = f\left(\beta x(s-1) + \rho u(s) + \Phi_0 \right). \label{eq:dde_ikeda_normalized}
\end{equation}
\noindent $x(s)$ during the normalized time interval of $s\in [0;1[$ can be regarded as the collective state, i.e. each delay interval contains a state-\emph{snapshot}.
Therefore, the relation between DDE-solutions and the virtual space-time representation is
\begin{equation}
x_{\sigma}(n)=x(s),\:n+\sigma=s,\:n=0,1,2,\ldots,T,\label{eq:space_time-1}
\end{equation}
\noindent where $\sigma\in[0,1[$ is a virtual space and $n$ is a new discrete time variable.
Under such substitution the DDE has a discrete time evolution along $n$ of a functional trajectory defined over time interval $\sigma$.

\begin{figure}[t]
	\begin{centering}
		\includegraphics[width=8.5cm]{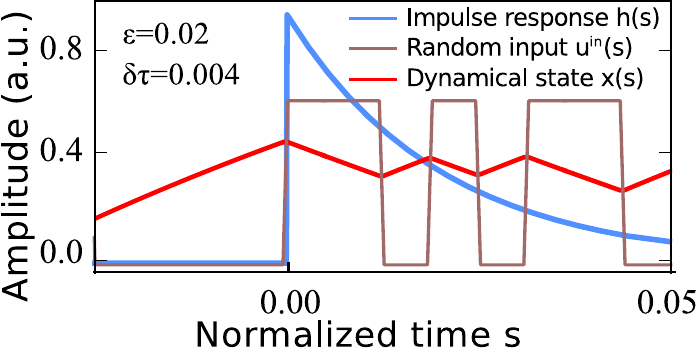}
		\par\end{centering}
	\caption{Impulse response function $h(t)$ (blue line), random binary input modulation $u^{int}(t)$ and and illustrative response of a low-pass Ikeda system $x(t)$ ($\varepsilon=0.02$, ), all in normalized time $s$.
		The sample-and-hold time for the random input sequence was
		Values located within a separation smaller than $h(t)$'s decay are coupled through convolution by, see Eq. \eqref{eq:functional_G}.
		For example, modulation by a random binary sequence (brown line) with a sample-and-hold time $\delta\tau={\varepsilon} / {3.2}$ continuously keeps the system's response (red line) from reaching a steady state.
		Consequently, responses to consecutive values in $u^{in}(t)$ are coupled through $h(t)$.}
	\label{fig:InertiaCoupling}
\end{figure}

While this transformation translates temporal positions $\sigma$ into a \emph{virtual space}, it does not reveal the network's nonlinear nodes or its connectivity structure.
Towards identifying these essential characteristics, we analyze the implications of the system's response to a delta perturbation, i.e. its impulse response function $h(s)$. 
For the low-pass system in Eq. \eqref{eq:dde_ikeda_normalized}, the response is
\begin{equation}
h(t) = \left\{ \begin{array}{ll}
e^{-t / T_R} / T_R; & t\geq 0  \\ \label{e18}
\qquad  0; & t< 0. \\
\end{array} \right.
\end{equation}
Graphically, Eq. \eqref{e18} is illustrated by a blue curve in Fig.~\ref{fig:InertiaCoupling}.
With the help of $h(t)$, Eq. \eqref{eq:dde_ikeda_normalized} can be rewritten as a global convolution product between impulse response function and the nonlinear function $f(\dots)$:
\begin{equation}
x(s)=\int_{-\infty}^{s}h(s-\xi)\cdot f\left(x(\xi-1)\right)d\xi. \label{eq:convolution_product}
\end{equation}
Using identity $s=n + \sigma$ in Eq. \eqref{eq:convolution_product}, we obtain
\begin{equation}
x_{\sigma}(n) = \int_{-\infty}^{n+\sigma} h\left(n+\sigma-\xi\right)\cdot f\left(x(\xi-1)\right)d\xi. \label{eq:x_sigma-iterated}
\end{equation}
\noindent The temporal dynamics of DDE Eq.~\eqref{eq:dde_ikeda_normalized} can therefore be interpreted as a functional sequence $G_{n},\:n=1,2,\ldots, T$, where $G_{n}:g_{n-1}\rightarrow g_{n}$ is a nonlinear integral operator mapping a function onto itself:
\begin{equation}
x_{\sigma}(n)=A_{\sigma}(n) + I_{\sigma}(n),\text{\ensuremath{\sigma\in}[0,1[}\label{eq:functional_G}
\end{equation}
\noindent where $A_{\sigma}(n)$ can be obtained via Eq. \eqref{eq:x_sigma-iterated},
\begin{equation}\
A_{\sigma}(n)=\int_{-\infty}^{(n-1)+\sigma}h\left(s-\xi\right)\cdot f\left[x(\xi-1)\right]d\xi . \\ \label{eq:convolution_product_outside_delay}
\end{equation}
Typical for delay-systems employed for RC is a configuration where $\varepsilon \ll 1$.
Within the integration-limits of Eq. (\ref{eq:convolution_product_outside_delay}) one can therefore assume the system's response to be negligible, leading to $A_{\sigma}(n) \approx 0$.
In the same manner, $I_{\sigma}(n)$ is obtained according to
\begin{equation}
I_{\sigma}(n)=\int_{(n-1)+\sigma}^{n+\sigma}h\left(s-\xi\right)\cdot f\left[x(\xi-1)\right]d\xi . \\
\end{equation}
Substituting variable $\xi'=\xi-n$, integral $I_{\sigma}(n)$ obtains the simpler form of
\begin{equation}
I_{\sigma}(n)=\int_{\sigma-1}^{\sigma}h\left(\sigma-\xi'\right)\cdot f\left[x_{\xi'}(n-1)\right]d\xi'.\label{eq:cont-time-network}
\end{equation}

Equation (\ref{eq:functional_G}) has the structure of a dynamical system being distributed along spatial dimension $\sigma$ and evolving with discrete time $n$.
Consequently, $x_{\sigma}(n-1)$ corresponds to a recurrency in time much like in a RNN, and $I_{\sigma}$ to coupling along virtual space $\sigma$ weighted by the impulse response function's amplitude at node-distance $\sigma-\xi^\prime$.
However, network state $x_{\sigma}(n)$ is a time-continuous signal within the interval $\sigma\in[0,1]$, while a networks' constituents are discrete nodes.
Yet, according to the Nyquist–Shannon sampling theorem, continuous signals can be discretized using a sampling interval less or equal to half the signal's fastest timescale.
For an autonomous DDE, continuous sequence $\sigma\in[0,1]$ then becomes a set of discrete values temporally separated by $\delta\tau\leq T_R / 2$.
$\delta\tau$ is the temporal separation between discrete states along virtual space $\sigma$, and node $l$ is therefore located at $\sigma_l = l\cdot\delta\tau$.
Combined, $\delta\tau$ and $h(t)$ address the discrete nodes' positions and their coupling, respectively.
Combined with Eq. \eqref{eq:x_sigma-iterated}, they provide the translation of a time-continuous DDE into a recurrent network of $N={\tau_D} / {\delta\tau}$ discrete nodes with nonlinearity $f(\cdot)$.

Yet, RNN's are not autonomous, and an external drive can exert considerable impact upon discrete system's symmetries introduced by $\tau_D$ and $\delta\tau$.
In Fig. \ref{fig:InertiaCoupling}, a nonlinear oscillator as in Eq. \eqref{eq:dde_ikeda} is modulated by signal $u^{in}(t)$ (brown line), in this case by a random Boolean sequence whose temporal separation between samples is significantly smaller than $\varepsilon=T_R / \tau_D$.
We therefore cannot rely on $T_R$ and the Nyquist–Shannon to define $\delta\tau$, and typically $\delta\tau$ corresponds to the shortest timescales of the injected signal.
Temporal aspects of $u^{in}(t)$ define $\delta\tau$ in most cases, also for $\delta\tau > T_R$.
As $u^{in}(t)$ corresponds to the information input into our recurrent delay network, it corresponds to the process which links information input to connection between virtual space $\sigma(t)$ and node $l(t)$.

\subsubsection{Time multiplexing: virtual photonic neurons}
\label{sec:DelayANNs_TempMultiplexing}

\begin{figure*}[!t]
	\begin{centering}
		\includegraphics[width=17cm]{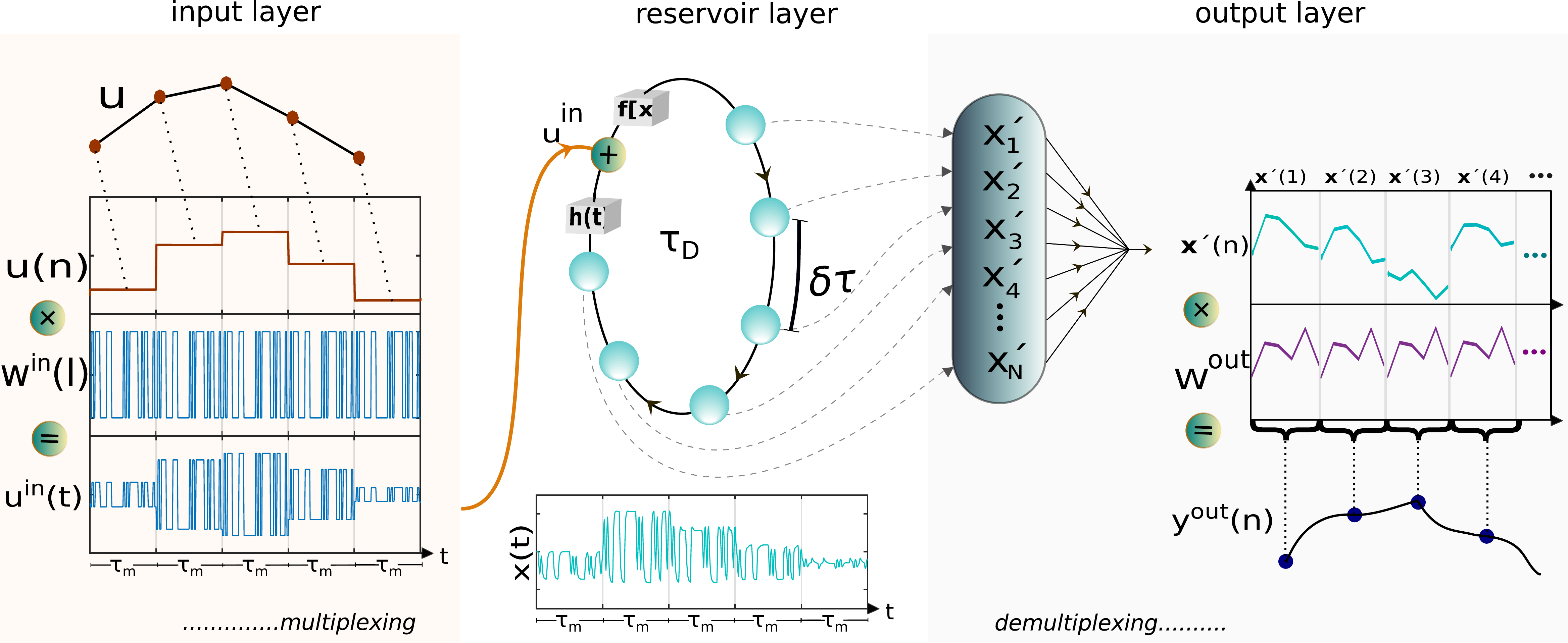}
		\par\end{centering}
	\caption{Schematic illustration of the individual processes involved in a delay-reservoir computer.
		The input layer is implemented by modulating input $u(n)$ with temporal mask $W^{in}(l)$ to create input $u^{in}(t)$, $l\delta\tau$ is the temporal position of node $l$ during one delay.
		The reservoir's response autonomously creates state $\mathbf{x}(n)$ during one delay.
		Readout weights correspond to temporal modulations of $\mathbf{x}(n)$ according to $W^{out}$, and summing the resulting sequence over the delay-length creates output $y^{out}(n)$.}
	\label{fig:rc}
\end{figure*}

A delay-based reservoir computing architecture can be summarized as follows.
An initial temporal masking distributes information along a temporal window typically close to delay time $\tau_D$.
This masking step addresses the chain of $N$ nonlinear oscillators as discussed in the previous section, which in turn nonlinearly transform the input information.
The computational result is produced by the output layer, where again information during a temporal window similar to $\tau_D$ is weighted by a temporal masking according to the readout weights $W^{out}$.
These three processes, schematically illustrated in Fig. \ref{fig:rc}, are conceptually comparable to classical RNNs.
Particular to a delay-based reservoir is that, instead of each neuron corresponding to an individual nonlinear node, a single physical nonlinear node provides all nonlinear transformations.

Temporal input-masking needs to randomly map the input information onto the delay Reservoir's temporal dimensions according to $W^{in}$.
This is generally achieved by a temporal encoding technique called \textit{time multiplexing}, with time multiplexing according to a simple input masking sequence visualized in the left panel of Fig. \ref{fig:rc}.
Entries of binary mask $W^{in}$ are multiplied onto a scalar input point $u(n)$, resulting in $N$-dimensional vector $\textbf{u}^{in}=(u^{in}_1, u^{in}_2, \dots, u^{in}_N)$ at each $n$, where $n\in\{1,2,\cdots,T\}$ is integer time.
Time multiplexing now assigns each entry of $\textbf{u}^{in}(n)$ to a temporal position given by $l\cdot\delta\tau,l\in\{1,2,\cdots,N\}$.
$l$ is therefore a particular virtual delay neuron's index, $\delta\tau$ their temporal separation.
The temporal duration to complete the input masking for one $n$ is mask duration $\tau_M = N\cdot\tau_D$.
Injecting this input-sequence into the delay system creates the essential nonlinear and high-dimensional transformation, whose details certainly depend on the delay-node's particular physical substrate.
This process is repeated for each $n$, and as such is the multiplication by $W^{in}$ and time multiplexing according to $\delta\tau$.
A delay-node's input typically consists of consecutive and non-overlapping sequences of length $\tau_m$.

Time multiplexing therefore links the input to the reservoir layer.
Linking the reservoir to the readout layer naturally requires a \emph{time de-multiplexing} step. 
As illustrated by $x(t)$ in the reservoir layer panel in Fig. \ref{fig:rc}, the delay-RNN's state so far only exists as time continuous signal $x(t)$.
Consequently, the first operation for time de-multiplexing is to divide the delay-system's continuous output $x(t)$ into non-overlapping intervals of duration $(n-1)\tau_m \leq t < n\tau_m$.
During each of these $\tau_m$ intervals, one assigns node values $x'_{l}(n)$ to the information contained in $x(t)$ during temporal intervals $t\in[(n-1)\tau_m + (l-1)\delta\tau , (n-1)\tau_m + l\delta\tau[$ for $l\in\{1,2,\cdots,N\}$.
This step de-multiplexes the signal by again assigning a temporal position to a virtual neuron and creating the RNN's state-vector. 
Once this state vector is obtained, one can implement training the delay RNN according to Sec. (\ref{sec:RC_training}) and the delay RNN's output is $y^{out}(n)= W^{out}\cdot \mathbf{x'}(n)$.

While time multiplexing and de-multiplexing provides the essential linkage between temporal positions and nodes, it also creates the essential ingredient of a RNN, the hidden layer's internal connectivity.
Two mechanisms have been demonstrated.
The first uses the system's infinite impulse response function.
The other a de-synchronization between delay and input mask duration $\tau_D$ and $\tau_m$, respectively.

The first demonstration of RC with a delay system \cite{Appeltant2011} uses the concept of dynamical coupling via impulse response function $h(t)$.
For this, the temporal duration of a single neuron $\delta\tau$ is kept shorter than the system's response time $T_R$.
In Fig. \ref{fig:InertiaCoupling}, the delay system's response (red data) was calculated for $\delta\tau=0.2 \cdot T_R$, keeping the system constantly from reaching its steady state as the broad $h(t)$ convolves values of adjacent nodes, thereby introducing the network's internal coupling.
Inertia-based coupling therefore results in the convolution-response between $h(t)$ and the neighboring nodes.
Hence, its extent is tunable via the ratio between $\delta\tau$ and $T_R$, which modifies the 'spatial' decay over positions $l\delta\tau$.
Advantageous of this concept is that it maximizes the delay RNN's speed and enables higher complexity internal connectivity via leveraging non-trivial impulse response functions.

The second approach was first demonstrated by Duport \textit{et al.} \cite{Duport2012}, increasing $\delta\tau$ significantly beyond $T_R$ and therefore creating a systems with an effectively instantaneous response.
A local coupling is introduced by $\delta\tau={\tau_D} / ({N+k})$, hence $\tau_m < \tau_D$.
The consequence of this mismatch is that node $x_l(n)$ is delay coupled to node $x_{l-k}(n-1)$.
The advantage of this concept is that the system closely resembles the nonlinear map of RNNs, which generally simplifies the mathematical model and a numerical simulation.
The reduced operational bandwidth can potentially be beneficial for the system's signal to noise ratio.
Importantly, both approaches are not mutually exclusive.
One can therefore create more complex network structures by merging both concepts.

Finally, we would like to touch upon some technical details of the time multiplexing implementation of RNNs.
If input $\textbf{u}^{in}(n)$ is calculated off-line, one can carry out the masking for each $u(n)$ and concatenate the $T$ resulting column-vectors $\textbf{u}^{in}$ to the full input matrix $\textbf{u}^{in}, \in\mathbb{R}^{N\times T}$.
The same is true for the recording the delay system's output $x(t)$: the de-multiplexing can be done off-line and in one batch for the entire data.
This has significant benefits for proof of concept experiments.
As generally the case of ANNs, the input can be an $M$-dimensional feature vector $\textbf{u}^m=(u^m_1, u^m_2, \dots, u^m_M)$.
Then, the result of masking by $W^{in}$, vector $\textbf{u}^{in}=(u^{in}_1, u^{in}_2, \dots, u^{in}_N)$, $N > M$ is a random weighting of the features from $\textbf{u}^{m}$.
Also, the input mask $W^{in}$ values is not restricted to binary entires.
And finally, $x_{l}(n)$ might be defined as the average or some specific value, for example at the middle or the end of discussed interval.

\subsubsection{Advantages of delay implementations of RNN}
\label{sec:DelayANNs_advantage}

Due to the recent success of ANNs, the scientific community needs explorer avenues for next generation implementations of such computational concepts.
Even simple neuromorphic architectures might conceptually be straight forward, yet they often are an exceptional challenge to be mapped directly onto corresponding hardware.
Spatial neural networks, as an example, comprise hundreds or even thousands of neurons, each connected to numerous other neurons in the network \cite{Bueno2018}.
Consequently, their hardware implementation require significant early stage efforts only for establishing the basic nonlinear network.
Accordingly, in order to process information efficiently, alternative routes to emulate neural network architectures are highly attractive \cite{Rodan2011}.

Due to the spatio-temporal analogy introduced in the previous sections, delayed feedback systems offer a highly attractive solution.
In particular combined with the advantages of photonic delay-lines, they are for now one of the few well-controllable and efficient experimental demonstration of large scale ANNs in physical substrates.
While these systems are of significant value for the demonstration of novel information processing concepts, they simultaneously allow for a detailed investigation of the underlying relevant processes.
Here we would like to highlight a feature unique to nonlinear delay systems: all nonlinear transformations carried out by the virtual spatio-temporal network rely on the same physical component.
In many cases it is therefore straight forward to obtain an accurate characterization of this component, which creates an excellent situation for comparing experimental and numerical/theoretical findings.
In this aspects delay system's occupy a strategic position in the field's future development.

\section{Photonic time delay Reservoirs}
\label{sec:PhotDelRRC}

The implementation of RC into physical substrates was quickly considered after the publication of the original concept.
Due to the fundamental spatio-temporal nature of a Reservoir, first considerations equally attempted to explore spatial dimensions for the implementation of the RNN.
First was a demonstration based on waves in a water tank \cite{Fernando2003}, which however did not surpass the processing performance beyond the linear system limit.
This was followed by the numerical demonstration in a network of multiple semiconductor optical amplifiers \cite{Vandoorne2008}.
However, while progress in such spatio-temporal systems was slow, RC based on delay systems quickly took off.

\subsection{First electronic proof of concept}
\label{sec:PhotDelRRC_ElectrDemonst}

The first implementation of a delay reservoir was realized in an electronic circuit by Appletant \textit{et al.} in 2011 \cite{Appeltant2011}.
In this seminal work, the authors demonstrated the fundamental operational concept based on the temporal multiplexing routine introduced in Sect. (\ref{sec:DelayANNs_TempMultiplexing}).
They investigated the impact of the temporal duration of virtual nodes $\delta\tau$, and findings from numerical simulations were used as guideline in the electronic system's design which consequently facilitated a tunable $T_R$ through RC-circuits. 
Several tasks have been successfully demonstrated, such as spoken digits recognition with NIST TI-46 corpus dataset, NARMA10, and Santa Fe time series prediction.
This work can be considered as the starting points for RC implementations in analog delay systems, and physical substrates in general.
However, in this tutorial we focus on the photonic implementations, which quickly followed.

\subsection{Electro-optical delay reservoirs}
\label{sec:EO_reservoirs}

\begin{figure}[t]
	\begin{centering}
		\includegraphics[width=8.5cm]{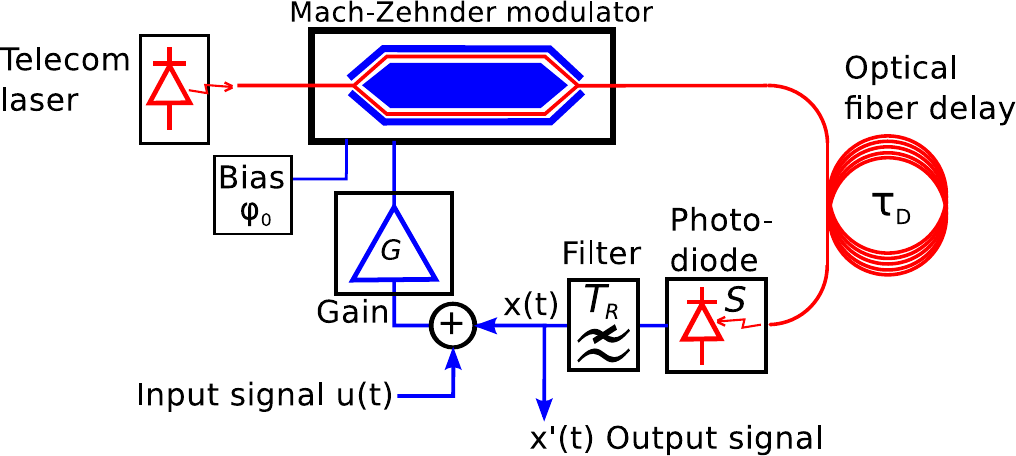}
		\par\end{centering}
	\caption{Electro-optical delay system RC \cite{Larger2012}.
		The system is based on telecommunication components, and the long time delay is realized via an optical fiber spool.
		Multiple systems based on similar architectures have been realized, including ultra-fast implementation reaching 20 GSamples/s data injection rates \cite{Larger2017}.}
	\label{fig:ikeda_RC}
\end{figure}

The first demonstrations of RC based on photonic systems \cite{Larger2012,Paquot2012} were realized electro-optically.
Both used comparable hardware setups, schematically illustrated in Fig. \ref{fig:ikeda_RC}.
While punctually modified, each electro-optical delay oscillator was based on earlier experiments in the field of nonlinear Ikeda dynamics \cite{Larger2004}.
A fiber pig-tailed semiconductor laser, here operated around telecom wavelengths ($\lambda\approx$1550~nm), was connected to a Mach-Zehnder modulator (MZM) creating a $\sin^{2}$ nonlinearity, whose output was injected into a delay line created by a long single mode optical fiber.
The delayed optical signal was detected by a photodiode, whose output was merged with an external drive.
The combined signals were injected into the MZM's radio-frequency port, thereby closing the feedback loop.
Depending on the details of the particular implementation, multiple amplifiers can condition signals at various stages.
Finally, filters can be added at different positions in the electronic section of the scheme, allowing the implementation of low \cite{Larger2004} or bandpass filtered dynamics \cite{Udaltsov2002}.
Without dedicated filters, response time $T_R$ depends on the slowest component of the electronic circuit.

For the here employed low-pass filtered system, the resulting DDE is 
\begin{equation} \label{eq:IkedaRC}
\varepsilon \dot{x}(s) + x(s) = \beta \sin^2 [ \mu x(s-1) + \rho u(s) +\Phi_0 ].
\end{equation}
Parameters and normalizations are as introduced in Secs. (\ref{sec:DelayANNs}) and (\ref{sec:DelayANNs_SpaceTempAnalogy}), respectively, with $\Phi_0$ is a typically constant DC phase-offset.
Convenient to implementations as the one shown in Fig. \ref{fig:ikeda_RC} is that each parameter of Eq. \eqref{eq:IkedaRC} is accurately controlled by a dedicated component.
Feedback gain $\beta$  either via the laser's intensity \cite{Larger2012} or via an optical attenuator placed inside the optical feedback line \cite{Paquot2012}, phase offset $\Phi_0$ simply via the voltage applied to the MZM's DC-electrode.
The system's response time $T_R$ can easily be characterized in an open-loop configurations, where the delayed feedback is simply disconnected.
This modularity allows for a highly flexible experimental system and enables access to most or even all relevant system parameters via detailed characterization.
Larger \textit{et al.} used fast modulation with $\delta\tau=$42.18~ns, which combined with $T_R=240$~ns resulted in $\delta\tau\sim{0.18}\cdot{T_R}$.
As introduced in Sec. (\ref{sec:DelayANNs_TempMultiplexing}), reservoir internal connectivity was therefore established via the system's inertia \cite{Larger2012}, and $\tau_D=$20.87$~\mu$s resulted in a reservoir with $N=$400 virtual neurons.

The system's capacity to serve as a photonic RC was first evaluated based on the TI46 spoken digit recognition benchmark test.
Spoken digit recognition corresponds to grouping input data into labeled classes, a computational task which does not necessarily require working memory.
From the overall Ti46 corpus, the authors selected 500 samples originating from five different female speakers uttering the digits from 0 to 9 with 10 repetitions.
The training target was to identify the number the spoken digit represents.
Readout weights $W^{out}$ were calculated using ridge regression based on the photonic reservoir's response to 475 randomly selected training samples.
The word error rate (WER) was evaluated based on the remaining 25 test samples which were not involved in the optimization of $W^{out}$.
Cross-validation ensured that training and test samples were iterated such that each sample was part of the testing set exactly once.
The system's WER dependence on dynamical parameters $\beta$ and $\Phi_0$ was characterized in detail, revealing that classification accuracy sensitively depends on $\Phi_0$, with best performance always located close but not exactly at the nonlinear
function's local extrema.
Bifurcation parameter $\beta$ proofed to be less delicate, and comparable performance was achieved for the range of $\beta \in [0.3; 0.6]$.
The lowest error found was WER = 0.4$~\%$, corresponding to only 2 wrong classifications out of the 500 spoken digits.

To evaluate the performance when utilized for solving a task explicitly requiring short term memory, the authors predicted the evolution of a chaotic signal.
The used Santa Fe prediction challenge is a well established prediction benchmark test, which requires the one-step ahead prediction of a chaotic infra-red laser's emission.
The training target therefore is the input information shifted by one time step into the future.
As before, readout weights $W^{out}$ were trained using 75$~\%$ of the data set, while prediction performance was evaluated based on the remaining 25$~\%$ data points of the test set.
Optimizing $\Phi_0$, the authors found that for this task, performance was best for phase offsets closer to the nonlinear function's inflection point, where the system's response is more linear.
Under such condition, the linear memory capacity is larger than for operation around a local extrema, where strong folding and stretching by the nonlinear function quickly quenches the system's linear echos.
Using the normalized mean square error as evaluation criteria, the lowest prediction error was a deviation of $NMSE=0.124$ between reservoir output and the correct future value, found at $\Phi_0=0.1\pi$ and $\beta=0.2$.

Paquot \textit{et al.} \cite{Paquot2012} employed the alternative approach of a temporal de-synchronization between mask ($\tau_m$) and delay time ($\tau_D$) to introduce reservoir internal connectivity, see Sec. (\ref{sec:DelayANNs_TempMultiplexing}).
With a delay system creating $\tau_D=$8.504$~\mu$s, they implemented a photonic reservoir of $N=$50 nodes by using a temporal mask duration of $T_R < \delta\tau=170~$ns.
In this case, the de-synchronization between mask length and $\tau_D$ corresponded to the duration of one virtual neuron $\delta\tau$, therefore $k=1$.

As a fundamental characterization, the authors determined the system's linear (31.9), quadratic (4) and cross (27.3) memory capacity, resulting in a total memory capacity of 48.6.
The first task addressed by Paquot \textit{et al.} is to reproduce a Nonlinear Auto Regressive Moving Average equation of order 10, driven by white noise (NARMA10) \cite{Paquot2012}.
Using the NMSE between target and photonic reservoir output, they obtain $NMSE = 0.1686 \pm 0.015$ both, in their experiment and in a numerical model of the same system.
This is an excellent results as crucially it is comparable to digital reservoirs of the same size, despite the noise inherent to analogue physical substrates and experiments in general.
As a second task, the authors considered the nonlinear channel equalization task already addressed in the work by Jaeger \textit{et al.} \cite{Jaeger2004}.
This task is of practical relevance, as it is a problem regularly encountered in wireless data transmission: reflections off various objects (e.g. buildings) arrive after different temporal delays and nonlinear transformations at the receiver, resulting in a deterioration of the recorded signal.
The task is then to reconstruct the original message from the recorded signal.
At an original signal to noise ratio of 28 dB, after being processed by the photonic reservoir the authors reported a symbol error rate of $SER=1.4\cdot 10^{-4}$, which again matches those of comparable digital emulations.
Finally, the authors too addressed the Ti46 spoken digit recognition task.
Using an enlarged reservoir of now $N=200$ virtual neurons, they report a $WER=0.4\%$ as best obtained performance, which matches the one reported by Larger \textit{et al.}.

Finally, the bandwidth of the electro-optical delay RC was recently pushed to the full potential of telecommunication components \cite{Larger2017}.
The setup was based on phase instead of intensity modulation, where before detection differential phase shift keying (phase shift $\delta_T=402.7~$ps) converts phase shifts into amplitude modulations.
A delay of $\tau_D=63.33~$ns and a virtual neuron duration of $\delta\tau=56.9~$ps results in $N=$1113.
A new data injection scheme referred to extended delay memory (EDM) too was introduced.
There, one $\tau_D$ is divided into $N_L$ sub-reservoirs, each of which operated according to the same procedures introduced before.
Consequence of this division is that input information $\mathbf{u}(n)$ mixes with $\mathbf{u}(n+N_L)$.
Furthermore, the authors implement a modified readout routine, applying a temporal spacing between readout samples of $\delta\tau^r=(1+\xi)\delta\tau$.
Input and readout clock are therefore de-synchronized, and it is shown that de-synchronization parameter $\xi$ has significant impact upon the system's performance.
As task, the authors addressed spoken digit recognition, this time under more challenging conditions than for the Ti46 data set.
To emulate real-world environments, in the AURORA-2 spoken digit benchmark voice recordings contain both sexes and include adults and children.
For $\beta=0.7$, $\Phi_0=2\pi / 5$, $N=1000$, $N_L=3$ and $\xi=5\cdot10^{-4}$, the authors report a $WER=0.4~\%$.
Most impressively, this ultra-high bandwidth implementation allows for a digit classification of 1 Million digits per second with such a simple device.

Electro-optical delay systems have paved the way for photonic RC and have by now been employed in multiple configurations addressing a wide range of benchmark problems with excellent performance.
Due to the excellent access to system parameters, they typically agree very well with numerical models.
Finally, they profit from highly developed off-the-shelf telecommunication industry components, making them equally attractive for first proof of principle experiments as well as for high-performance implementations.

\subsection{Electro-optical reservoirs and extreme learning machines}
\label{sec:EO_RC_ELM}

An interesting implementation option arises from the general structure of delay systems.
A radically simplistic ANN concept exclusively uses the mapping of input information onto individual, uncoupled nonlinear nodes.
The concept, referred to as extreme learning machines (ELM) \cite{Guang-BinHuang2004}, therefore does not require internal connections between neurons.
In delay systems, this corresponds to setting $\delta\tau > T_R$ and to disconnecting the delay.
One can therefore conveniently alternate between operating the same photonic system as a delay RC and a photonic ELM, simple by modifying a single hardware connection.
Such a photonic ELM and its comparison to a photonic delay RC was implemented based on the first introduced electro-optical experimental setup \cite{Larger2012}.
Hence, this experiment demonstrated a unified framework for RC and ELMs based on identical hardware \cite{Ortin2015}.
Results obtained for the ELM were compared with the RC approach and confirmed via numerical models, including a compensation of inherent working memory loss via the data injection approach when the system was operated in the ELM configuration.

\subsection{All-optical Reservoirs}
\label{sec:AO_RC}

The transition from opto-electronic to all-optical RC is of fundamental importance for multiple reasons.
First, because such systems avoid conversions between optical and electronic signals.
These potentially result in additional sources of noise, bandwidth limitations and in an reduction of the overall energy efficiency.
Second, many sensors as well as information communication protocols operate with optical signals; those and all-optical ANNs could therefore be interfaced directly.
Based on numerical simulations, the earliest consideration of an all-optical RC suggested the connection of various semiconductor optical amplifiers (SOAs) into a simple spatio-temporal network \cite{Vandoorne2008}.
The architecture directly targeted integration into a photonic chip, and a integrated system having the same architecture yet avoiding the nonlinear SOAs was published in 2014 \cite{Vandoorne2014}.

\begin{figure}[t]
	\begin{centering}
		\includegraphics[width=8.5cm]{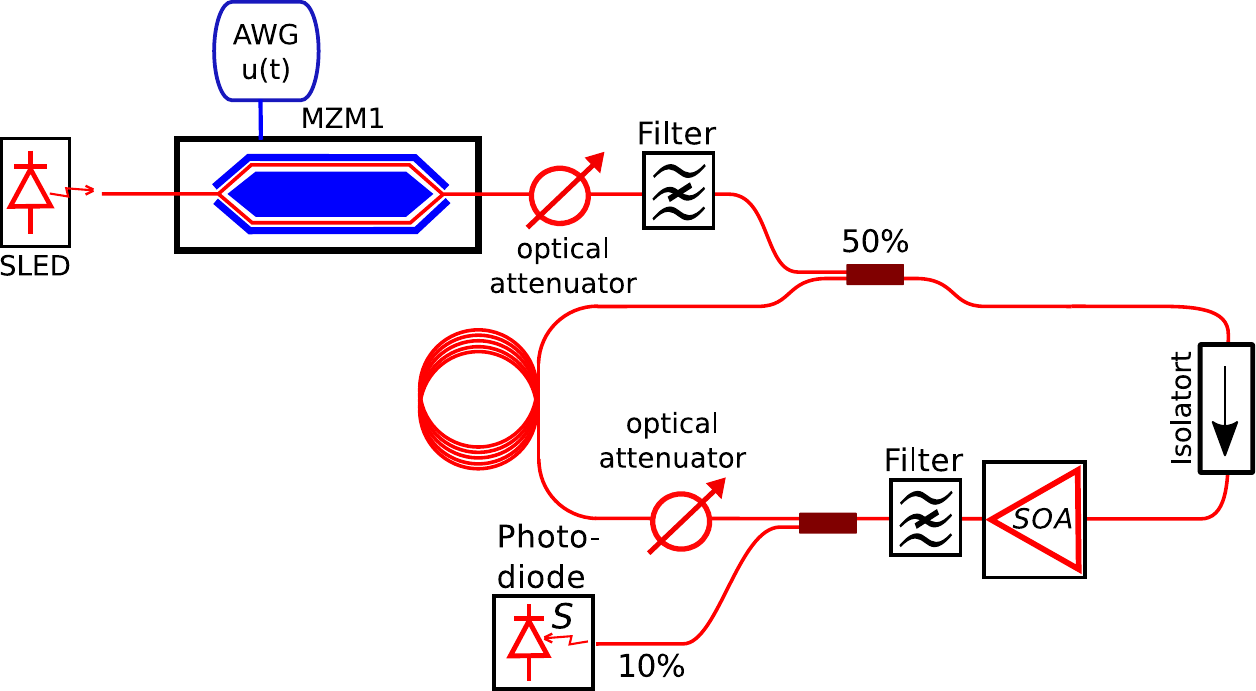}
		\par\end{centering}
	\caption{All-optical RC based on a  semiconductor optical amplifier (SOA), acting as all-optical nonlinearity, located within a long delay loop \cite{Duport2012}.
		The SOA is driven below the lasing threshold of the ring-cavity laser, and information is injected all optically.
		A delay time of $\tau_D=7.9437~\mu$s and $N=50$ virtual neurons resulted in a modulation bandwidth of 6~MHz.}
	\label{fig:SOA_delay_RC}
\end{figure}

The first all-optical demonstration of a Reservoir was based on a delay loop which included a semiconductor optical amplifier (SOA) acting as all-optical nonlinearity, see Fig. \ref{fig:SOA_delay_RC}.
When driven into saturation, these devices experience a nonlinearity which resembles a hyperbolic tangent.
This particular nonlinearity is very popular in ANNs and computational neuro-science, since it approximates the spiking rate of biological neurons.
Duport \textit{et al.} \cite{Duport2012} located the SOA within a long fiber optical delay loop with a delay of $\tau_D=7.9437~\mu$s.
In this configuration the SOA acts as amplification medium within a ring-cavity, and optical attenuation inside the cavity and SOA bias current were adjusted such that the ring laser was operated below its lasing threshold.
The authors implemented a reservoir of $N=50$ virtual nodes, employing de-synchronization of $k=1$ between mask and delay length $\tau_m$ and $\tau_D$, as introduced in Sec. (\ref{sec:DelayANNs_TempMultiplexing}).
The system was injected with the light of a semiconductor superluminescent diode, which was modulated via a MZM to encode the masked input information.
The authors evaluated the system's memory capacity, obtaining a linear and total memory capacity of 20.8 and 28.84, respectively, which is significantly below the one previously reported for their opto-electronic system.
Implementing the previously introduced channel equalization task, the authors obtained a symbol error rate of $\sim5.5\cdot10^{-4}$ for a signal to noise ration of 28 dB.

\begin{figure}[t]
	\begin{centering}
		\includegraphics[width=8.5cm]{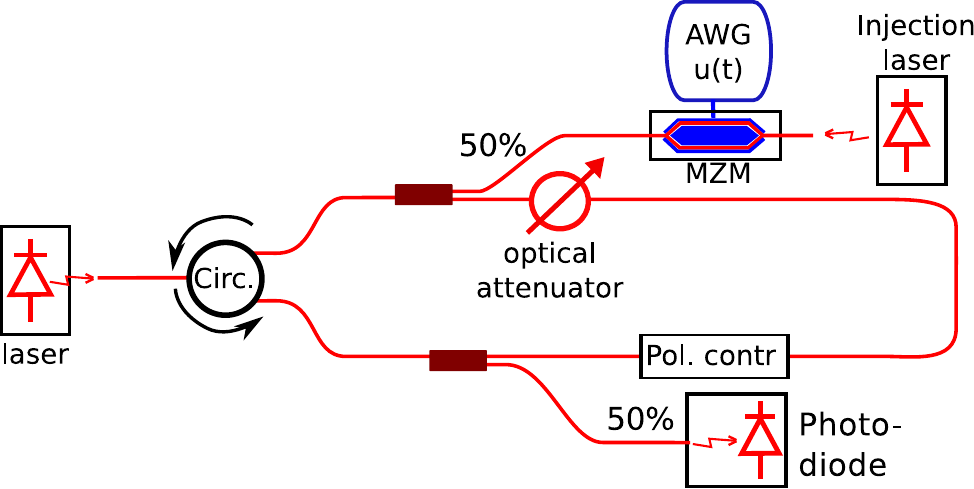}
		\par\end{centering}
	\caption{All-optical RC based on a  semiconductor laser diode, acting as all-optical nonlinearity, coupled to a long delay loop \cite{Brunner2013a}.
		A delay time of $\tau_D=77.6~$ns and $388$ virtual neurons resulted in a modulation bandwidth of 5~GHz.
		Information was injected, both, optically via an external injection laser, and electrically via modulating the laser's bias current.}
	\label{fig:Delay_laser_RC}
\end{figure}

Shortly after, another all-optical implementation was demonstrated by Brunner \textit{et al.} \cite{Brunner2013a}, where they realized the first ANN based on a semiconductor laser.
The particular appeal of semiconductor lasers ANNs has been discussed previously \cite{Hoppensteadt2000}.
Specifically, semiconductor lasers are energy efficient, high-bandwidth and react strongly nonlinear to optical injection \cite{Wieczorek2005}.
The experimental setup is schematically illustrated in Fig. \ref{fig:Delay_laser_RC}.
This form of all-optical Reservoir was efficiently implemented, and consisted only of the telecommunication laser diode, a fiber circulator (Circ.), a polarization controller (aligned to rotated feedback polarization), an optical attenuator and  two optical fiber splitter for signal injection and readout.
The fiber optical delay line had a round-trip delay of $\tau_d=77.6~$ns, and $N=388$ virtual neurons were implemented  using a 5 GHz modulation bandwidth.
Information injection was implemented all-optically via an external injection laser, as well as electrically via direct modulation of the laser's bias current.
For that, the laser was biased with a bias-tee, whose rf-port allowed high frequency current modulation.

Using the Ti46 spoken digit recognition benchmark as introduced in Sec. (\ref{sec:EO_reservoirs}), the authors compared both information injection techniques and their performance dependence on the laser's DC-bias current.
Electrical signal injection resulted in a word error rate of 0.64$~\%$, while for all optical operation the error dropped dramatically to 0.014$~\%$.
Simultaneously, the authors extended the test and utilized the same Reservoir responses to identify the speaker of the digit with an error rate of 0.88$~\%$.
At the data-rate used in this experiment, the recognition rate was 300,000 spoken digit per second.
Exploiting the recurrent nature of the system, the authors trained the same Reservoir for the prediction of the Santa Fe chaotic laser benchmark, where they achieved a prediction NMSE of 0.106.

This demonstration of all-optical RC based on a standard, off-the-shelf semiconductor laser inspired numerous follow up experiments.
The same group of authors demonstrated fast all-optical vector and matrix operations based on the same system \cite{Brunner2013} and later extended the computation performance analysis with a detailed study focusing on parameters crucial for drive-response laser systems \cite{Bueno2016}.
Finally, they published the numerical model of the experiment \cite{Hicke2013}.
A modification of the laser RC system was presented by Nguimdo \textit{et al.} \cite{Nguimdo2015}, where they used a semiconductor ring laser to solve two different tasks in parallel utilizing the clock and counter-clockwise propagating beams.
In a comparable system, the same group of authors showed that one can significantly reduce the system's sensitivity to phase drifts by combining the Reservoir's original output and an additional copy delayed by more than a delay time $\tau_D$  \cite{Nguimdo2016}.
The main authors also investigated RC based on a Erbium doped microlaser, both in experiments and numerical simulations \cite{Nguimdo2017}.

Finally, other all-optical implementations of RC were demonstrated on a passive resonator \cite{Vinckier2015} and a saturable absorber \cite{Dejonckheere2014}.
Particular to both systems is that they do not contain an optical gain-element and are therefore low noise.
The passive resonator RC demonstrated excellent performance in the nonlinear channel equalization task, while the authors found the saturable absorber RC to provide an excellent linear memory capacity of 36.8.

\subsection{Improvements to photonic delay RC}
\label{sec:EXP_RC_input_mask}

Multiple studies investigated possible strategies in order to improve the performance of the photonic delay RC.
Properties of the input mask can be seen as an intuitive starting point, as it is the first operation carried out on the data to be processed.
In the first delay RC demonstration, the originally uniformly and randomly distributed mask was replaced by a random Boolean sequence \cite{Appeltant2011}.
Soriano \textit{et al.} \cite{Soriano2013} investigated how increasing the number of masking values from 2 to 6 suppresses the impact of digitization noise.
Nakayama \textit{et al.} \cite{Nakayama2016} went one step further and analyzed the impact of complex masks.
They found that, for prediction, using a chaotic mask whose highest spectral density is located at the delay-laser's relaxation oscillation frequency $\nu_{RO}$ improves performance.
Finally, Appeltant \textit{et al.} \cite{Appeltant2013} constructed a Boolean mask based on the minimal length sequence technique, resulting in a reduced length and hence a reduced number of nodes resulting in a faster system update rate at same performance.

Another strategy to elevate performance of a photonic delay RC was to implement multiple delays.
This was first demonstrated by Martinenghi \textit{et al.} \cite{Martinenghi2012} based a electro-optical systems based on complex wavelength dynamics \cite{Larger2017}.
The authors added numerous delays of length shorter $\tau_D$ to the delay architecture.
For the reservoir of $N=150$ virtual neurons, they included $N_I=\dfrac{N}{10}$ additional delays of randomly selected  integer multiples of $\delta\tau$.
Besides the short range coupling via the impulse response function $h(s)$, the delay therefore adds additional internal connections between 10$~\%$ of the virtual neurons.
Recently, Hou \textit{et al.} \cite{Hou2018} have investigated the impact of adding a second delay which is  $\tau_{d}^{2}=\tau_{d}^{1}+\dfrac{1}{2\nu_{RO}}$ and $\tau_{d}^{1} = \tau_m + \delta\tau$, where $\tau_{d}^{1}$, $\tau_{d}^{2}$ are time delay of the first and second delay loop, respectively.
In their numerical simulations, they found that prediction performance for such a double delay configuration does not require fine tuning of $\delta\tau$ and generally shows improved performance.

\subsection{Electro-optical reservoirs with input and output weights}
\label{sec:EO_Input_Output}

\begin{figure}[!t]
	\begin{centering}
		\includegraphics[width=8.5cm]{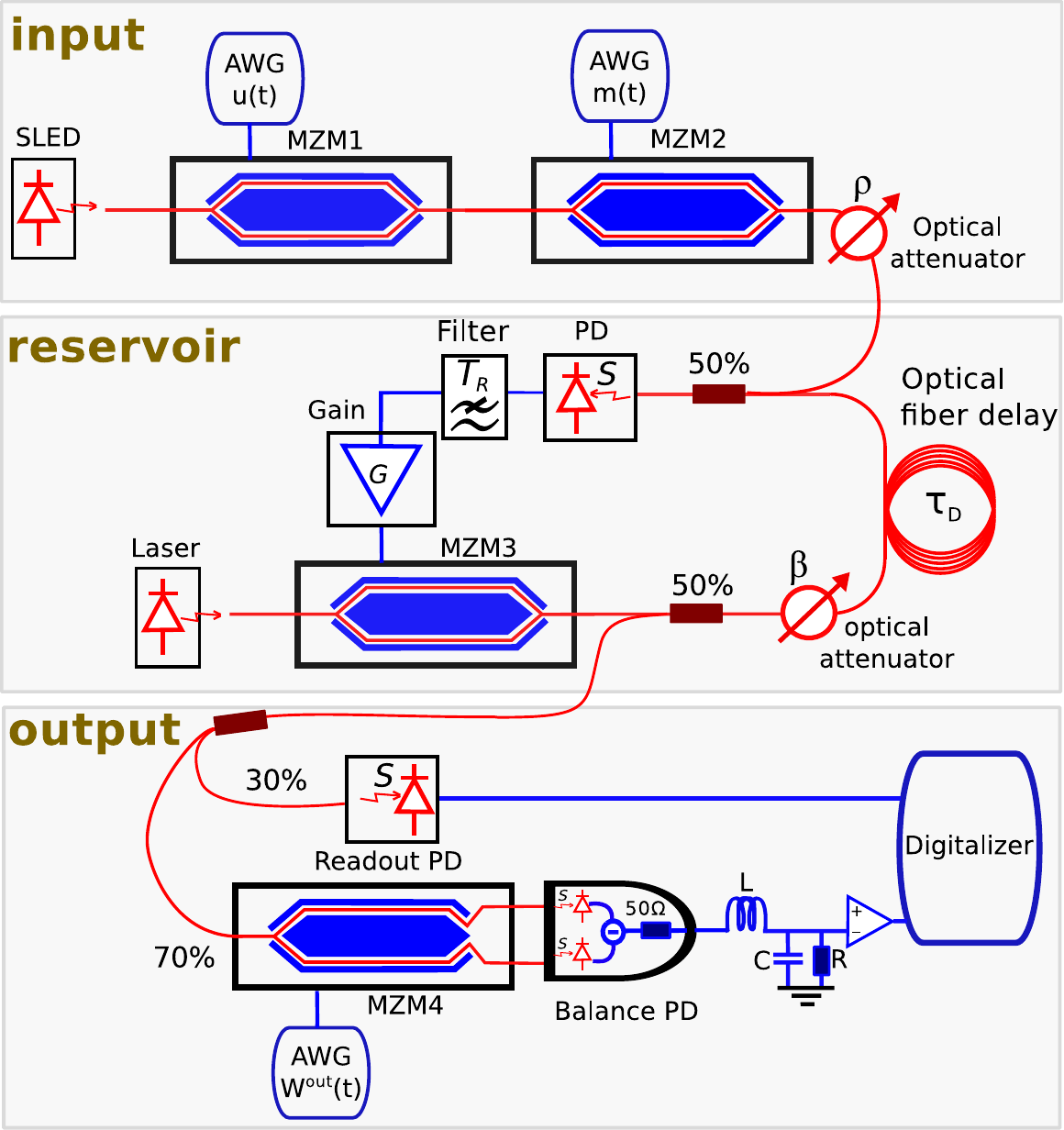}
		\par\end{centering}
	\caption{Implementation of all of a delay-reservoir computer's layers in photonic hardware \cite{Duport2016}.
		Starting from the input superluminescent diode, the first two MZMs realize the input signal $u(n)$ and its temporal masking according to $W^{in}(l)$, realizing signal $u^{in}(t)$.
		The following delay-reservoir is identical to \cite{Paquot2012}.
		Readout weights $W^{out}$ are based on the final, dual-output MZM, which is combined with a balanced-detection PD in order to allow for bipolar readout weights.}
	\label{fig:analogue}
\end{figure}

Until this stage, matrix multiplications according to input mask $W^{in}(t)$ and read out weights $W^{out}$ have been implemented off-line, typically after transferring data between the experiment and a standard desktop PC.
Such data preprocessing is common-sense for proof-of-concept experiments.
They correspond to simple linear multiplications and until now the focus was on the implementation and investigation of the physical Reservoir.
Yet, for creating a fully implemented and potentially standalone photonic RC, input mask and output weights need to be incorporated in hardware as well.
Such an analogue photonic neural network could significantly increase the system's energy efficiency, and removal of the off-line process will strongly speed up the operation as it enables operations of the full computing system in realtime.

The first demonstrations of an analog implementation of input masks \cite{Duport2014}, and finally of input as well as the system's output weights \cite{Duport2016} were both demonstrated based on the opto-electronic setup of Sec. (\ref{sec:EO_reservoirs}).
The experimental implementation of the experiment is schematically illustrated in Fig.~\ref{fig:analogue}.
Following the temporal multiplexing and de-multiplexing concept of Sec. (\ref{sec:DelayANNs_TempMultiplexing}), input mask $W^{in}$ and readout weights $W^{out}$ are temporal sequences to be linearly multiplied with input $u(n)$ and reservoir state $x(t)$, respectively.
As illustrated by Fig. \ref{fig:analogue}, input signal $u(n)$ is realized by modulating the optical intensity $I_{0}$ of a superluminiscent light emitting diode via intensity modulator $MZM1$, whose rf-input is connected to information input $u(t)$.
A second modulator ($MZM2$) follows and realizes the input masking, for which the authors demonstrated a novel strategy.
Instead of masking via a random sequence, the mask is defined by the addition of two sinusoidal functions at frequencies $p/\tau_{D}$, and $q/\tau_{D}$, with $p,q$ integers.
The resulting signal is nonlinearly mapped by $MZM2$, causing the final mask $W^{in}$ to be of higher complexity:
\begin{equation}
\begin{split}
W^{in}=\dfrac{1}{2} \{1 &+ \sin [ -\dfrac{\pi}{4}\cos (\dfrac{2\pi p}{\tau_{D}}t ) \\
&-\dfrac{\pi}{4}\cos (\dfrac{2\pi q}{\tau_{D}}t ) ] \}.
\end{split}
\end{equation}
\noindent While this strategy certainly  will not be able to create the high-dimensional mapping of a random input mask, it introduces an elegant strategy for fully-analog photonic implementation of a RC's input layer.
Finally, an optical attenuator adjust the intensity of the created input signal, corresponding to the Reservoir's input scaling $\rho$.
The reservoir layer itself follows the same scheme introduced in Sec.~\ref{sec:EO_reservoirs}.

Light transmitted to the output is now divided into $30\%$ to be detected by a readout photodiode, used for off-line training in order to calculate the output weights $W^{out}$.
The remaining $70\%$ feeds the dual output $MZM4$, which after the off-line training will be driven by $W^{out}$ provided by an AWG.
Both MZM output-ports provide complimentary signals which are detected by balance photodiodes, therefore allowing the implementation of bi-polar readout weights.
Combining a dual-port MZM and balanced detection is of significant importance, as the otherwise always positive optical signals would severely hinder system optimization via Eq. \eqref{eq:ridge-regression}, consequently strongly deteriorate the quality of computation.
Finally, creating the reservoirs output $y(t)$ requires integration of $x(t)$ weighted by $W^{out}(t)$ during one delay $\tau_D$.
Here, the authors resort to a simple filtering via a low-pass RLC filter.
This scheme has successfully demonstrated nonlinear channel equalization, the NARMA10 task and demonstrated  radar signal forecasting \cite{Duport2016}.

\subsection{Autonomous photonic delay Reservoirs via FPGA interfacing}

Finally, the demonstration of fully stand-alone photonic delay RC was realized using a field-programmable gate array (FPGA) as auxiliary infrastructure providing real-time control of the system.
The importance of this demonstration is two-fold.
First, it shows that real-time and high-bandwidth photonic ANNs are feasible.
Second, in many applications one is forced to consider the impact of a slowly changing environment modifying parameters of the physical RC as well as the target signal.
In order to compensate for these modifications, the readout weights may need to be updated online.
This requires a flexible standalone system.
A combination of analog-digital subsystems is used to facilitate this functionality, where the is Reservoir analog, while input and readout layers are implemented by the FPGA.
An FPGA device consists of electrical circuits (logical gates) that are interconnected using internal memory (lookup tables).
Programming FPGAs is achieved via setting the lookup tables' values, which	provides device rewiring without performing a manual interconnection.
With this technique, FPGAs can be set to perform data preprocessing, input masking, and linear readout.
Finally, readout weights can be adjusted online, i.e. on a running RC system.

Since FPGAs are essentially digital devices, they are interfaced to analog reservoirs via analog to digital and digital to analog converters.
The conjunction of FPGAs and analog optical reservoirs was successfully demonstrated by Antonik \textit{et al.} \cite{Antonik2016}.
In this work, a gradient descent algorithm was executed by an FPGA in order to update the readout values according to a certain target signal.
The reservoir was operated with the delay-mask de-synchronization, introduced in Sec. \ref{sec:DelayANNs_TempMultiplexing}.
This optoelectronic implementation has been successfully applied to the nonlinear channel equalization \cite{Antonik2016} and chaotic time series prediction \cite{Antonik2017} tasks.

\subsection{Back propagation through time}

\begin{figure}[t]
	\begin{centering}
		\includegraphics[width=8.5cm]{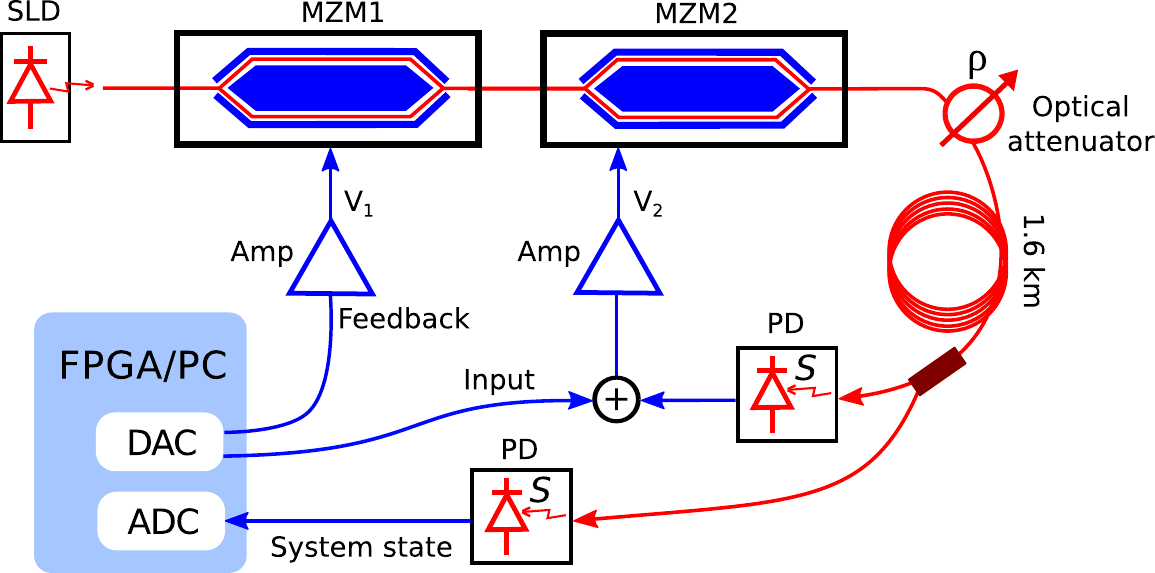}
		\par\end{centering}
	\caption{Backpropagation in hardware\cite{Hermans2016}:
		the light is emitted by a superluminescent diode (SLD) and 
		passes through two coupled Mach-Zender interferometers MZM1 and MZM2.
		The first interferometer is controlled by voltage $V_1$ (feedback signal) and the second one, by voltage $V_2$ (input signal).
	}
	\label{fig:backpropRC}
\end{figure}

Reservoir Computing is based on random feature extraction, hence a high dimensional state space is required to increase the probability of creating a successful reservoir dynamics \cite{Hermans2015b}.
In a single-node architecture, this may be achieved by increasing the delay time, which, however, reduces the system's computation speed.
A viable alternative to completely random connections is full optimization of the machine learning model.
A technique called backpropagation through time\cite{Hermans2010,Hermans2014} enables system optimization while at the same time keeping the reservoir's dimensionality $N$ relatively small.
The advantage of the proposed scheme is the ability to perform backpropagation directly in hardware, thus
resolving the need for computer simulations.
The technique is an extension of a common ANN training scheme \cite{Rumelhart1986} to continuous-time recurrent networks.

The general idea of the backpropagation method is first to run the system in forward direction and measure the error signal.
Then, the error gradient is propagated in reverse direction, i.e. starting with the output layer, taking into account the dynamical system memory effect.
In hardware that is achieved using a reciprocal system, i.e. reversing the roles of the receiver and transmitter, which is fed a time-reversed error signal.
Since the dynamical system is typically fixed in delay RC, the backpropagation may be used to optimize both the readout $W^{out}$ and the input mask $W^{in}$.

A first implementation of back propagation training in an optoelectronic system was demonstrated by Hermans \textit{et al.} \cite{Hermans2015}.
More recently, an experimental photonic demonstration \cite{Hermans2016} was based on an optoelectronic system consisting of a superluminescent diode, two coupled Mach-Zehnder modulators, a long spool of fiber as an optical delay line, interfaced with an FPGA to generate and record signals, and a computer controlling the experiment (Fig. \ref{fig:backpropRC}).
The system was configured such that the pair of Mach-Zehnder modulators was used either as a sine function in forward pass or as a cosine (reciprocal system) in the backward pass.
The evaluated benchmark tests were predictions tasks NARMA10 and VARDEL5 and TIMIT phoneme classification.
The authors showed that training both input mask and readout provides a better accuracy than just training the readout as in conventional RC.

\subsection{Beyond delay reservoirs}

The implementation of RNNs in photonic delay systems has had a profound impact upon the physical implementation of ANNs.
This is mostly due to the concept's elegance and simplicity when it comes to physically implementing the hidden recurrent layer.
However, there is no free lunch, and the simplicity is bought in expense of time multiplexing and de-multiplexing in the input and readout layer.
Both processes require accurate clocking plus the non-volatile storage of input and readout weights.
As a consequence, until now this functionality has only been realized either based on off-line pre and post-processing, or by heavy usage of an auxiliary infrastructure i.e. in form of a FPGA.
Furthermore, temporal multiplexing results in a reduction of the system's overall processing bandwidth by the number of neurons $N$.
Where for a fully implemented spatio-temporal network, the entire system's state would be available after $\delta\tau$, in delay systems this update requires $\tau_m=\delta\tau\cdot N$.
Plenty of promising and interesting approaches to implement ANNs in physical hardware substrates therefore remain to be explored.

\section{Outlook}

After the highly successful demonstrations of the past decade, photonic RNNs in delay system's will continue to enable both, technological applications as well as providing fundamental insight.
These directions of future development fundamentally profit from the unique properties of delay systems: an almost unbeatable experimental simplicity combined with the potentially high complexity resulting from their infinite dimensional phase space.
This will allow easy transfer of a technological application to the most recent and powerful nonlinear substrate.
An example is the successful experimental and numerical demonstration of channel equalization in short-range and long-haul optical communication systems \cite{Argyris2018}, another the detection of extreme-event \cite{Antonik2016b}.

At the same time, delay systems elegantly implement a perfectly symmetric ring-network, creating excellent conditions for investigating the interplay between fundamental properties of the photonic networks and their capability to process information \cite{Marquez2017,Marquez2018}.
These include the symmetry breaking via novel input masking or by more complex multiple delay architectures.
Simulateneously, the delay concept enables easy direct access to each virtual node for networks of almost arbitrary size, which could shift the focus to more local analysis of the photonic RNN \cite{Uchida2004}.

Implementation into novel substrates certainly will be an additional avenue of potentially important future development.
These include possibilities arising from ultra-high speed optical nonlinearities as well as novel photonic devices with superior energy efficiency \cite{Liu2015}.
Finally, explorations beyond the single delay node photonic RC architecture are of pressing importance.
These include investigation of multi-node Reservoirs \cite{Akrout2016}, implementations into large scale spatio-temporal networks \cite{Bueno2018} and into integrated photonic chips \cite{Shen2016,Katumba2018,Tait2017}.

\section{acknowledgments}

This work was supported by the Labex ACTION program (Contract No. ANR-11-LABX-0001-01), by the BiPhoProc ANR project (ANR-14-OHRI-0002-02), by the Volkswagen Foundation NeuroQNet project, the CNRS via the project PICS07300 and the Ministerio de Economía y Competitividad via project IDEA (TEC2016-80063-C3-1-R).

\bibliography{bibliography}

\end{document}